\documentclass[pdflatex,sn-mathphys-num]{sn-jnl}

\usepackage{graphicx}%
\usepackage{multirow}%
\usepackage{amsmath,amssymb,amsfonts}%
\usepackage{amsthm}%
\usepackage{mathrsfs}%
\usepackage[title]{appendix}%
\usepackage{xcolor}%
\usepackage{textcomp}%
\usepackage{manyfoot}%
\usepackage{booktabs}%
\usepackage{algorithm}%
\usepackage{algorithmicx}%
\usepackage{algpseudocode}%
\usepackage{listings}%

\begin{document}

\title{Improving YBa$_2$Cu$_3$O$_{7-\delta}$ annealing times through a combining-temperatures route}

\author*[1]{\fnm{Roberto F.} \sur{Luccas}}\email{luccas@ifir-conicet.gov.ar}
\author[2]{\fnm{Lorenzo} \sur{Gallo}}

\affil[1]{\orgdiv{Instituto de F\'{\i}sica Rosario}, \orgname{CONICET-UNR}, \orgaddress{\street{Bv. 27 de Febrero 210bis}, \city{Rosario}, \postcode{S2000EZP}, \country{Argentina}}}
\affil[2]{\orgname{Universidad Nacional de Rosario}, \orgaddress{\street{Pellegrini 1120}, \city{Rosario}, \postcode{S2000BTP}, \country{Argentina}}}

\abstract{The oxygenation process at constant temperature of YBa$_2$Cu$_3$O$_{7-\delta}$ (YBCO) was systematically investigated in the temperature range from 300 $^o$C to 800 $^o$C.
With this purpose, fully deoxygenated powder samples was exposed to an oxygen saturated atmosphere, and the evolution of their mass was recorded as a function of time using a thermogravimetric balance.
Results reveal a strong dependence of both: the oxygenation kinetics and the final oxygen saturation level, on the temperature used for oxygenation.
Moreover, results show that higher oxygen temperatures promote faster oxygen absorption but lead to lower saturation levels (higher final $\delta$ values), whereas lower oxygen temperatures result in slower kinetics but enable the system to approach better oxygenation conditions in order to improve the final superconducting properties of the material.
In addition to our measurements, a comparative analysis between oxygenation levels at the oxygen temperatures under study was performed in the range around oxygen saturation ($\delta$ $<$ 0.3).
Consequently, an oxygenation protocol based on a combination of several oxygenation temperatures is proposed.
As a first approach, results from a protocol with just two different oxygenation temperatures is compared with results coming from using just one oxygenation temperature.
Outstandingly, a protocol with a first oxygenation step at high temperatures and a second one at low temperatures demonstrates to improve oxygenation times in near a 30 \% for reaching $\delta$ values below 0.1 and in near a 60 \% for reaching $\delta$ values around 0.12.
Finally, we trust that our results are of direct application on industry since size of grain used herein are in scales of typical thickness of superconductor tapes.}


\maketitle

\section{Introduction}\label{SecIntro}

One of the main characteristics of the YBa$_2$Cu$_3$O$_{7-\delta}$ (YBCO) compound is the strong dependence of its superconducting and physical properties on the oxygen content $\delta$\cite{CupratesBM,Norton98,Wang24}.
The oxygen content determines both the crystal structure and the carrier density within the CuO$_2$ planes, thereby directly affecting key superconductor parameters such as critical temperature T$_C$, critical current density J$_C$ and irreversibility line H$_{irr}$\cite{Solovjov19,Chikumoto05,Mori10}.
In particular, when the material exhibits high $\delta$ values, it adopts a tetragonal structure lacking of superconducting properties, whereas at low $\delta$ values the orthorhombic phase prevails exhibiting a superconducting behavior\cite{Tu87}.

For this reason, the post-growth oxygenation process constitutes a critical step in the preparation of YBCO superconductor samples.
In many synthesis methods, like PLD, MOD-TFA or even the flux zone among others; the as-prepared material initially presents an oxygen content lower than that required to achieve functional superconducting properties\cite{Norton98,Chikumoto05}.
Consequently, post-annealing thermal treatments in oxygen atmospheres are necessary.
During this process, oxygen diffuses into the crystal lattice and occupies vacancies in the Cu-O chains, leading to the structural and electronic transition required for the establishment of the superconductivity\cite{Cui05}.

Although not yet fully understood, the mechanisms governing oxygen incorporation in YBCO involve a complex combination of processes that depend on the form in which the material is presented.
Basically, these include the inclusion of oxygen atoms in the crystal structure of the material at the surface of it (adsorption, chemical reaction, etc.) and the diffusion of them to the inner layers of the material (oxygen diffusion).
All of which exhibit kinetics strongly dependent on temperature, oxygen partial pressure, and the microstructure of the YBCO material\cite{Stangl25}.
Early studies demonstrated that oxygen diffusion in YBCO is highly anisotropic and that achieving stoichiometric equilibrium may require prolonged annealing times ($>$ 200 hours), particularly in dense or thick samples\cite{Tu87,Zheng03}.

Several studies have attempted to optimize this process by modifying thermal treatment parameters such as annealing temperature, oxygen pressure or treatment duration\cite{Chikumoto05,Mori10,Stangl25,Qu13}.
Although the results vary depending on the form of the material (thin films, single crystals, powder, tapes, etc.), there is agreement on several key aspects.
In general, oxygen mobility decreases at low temperatures, while higher temperatures enhance mobility with a reduction of oxygenation times but also lowering the final oxygen saturation values.
Additionally, it has been observed that the material incorporates oxygen more rapidly than it is deoxygenated\cite{HudnerPhDThesis}.

However, despite the numerous studies reported, the optimization of oxygenation processes remains veiled, despite its crucial role in achieving materials with optimal properties, particularly when different material presentations are considered\cite{Chikumoto05,Mori10,Stangl25,LuccasArXiv}.

In this context, the present work focuses on the experimental study and optimization of the oxygenation process in powdered YBCO.
Based on data obtained from oxygenation measurements performed at different temperatures, a temperature ramp combining multiple regimes is proposed to reduce oxygenation times while minimizing the final $\delta$ value achieved.
This work aims to determine the optimal thermal treatment conditions in oxygen atmospheres to enhance superconducting properties of the material while simultaneously reducing oxygenation times.

In conclusion, this study is intended to serve as a basis for the optimization of the oxygenation process in YBCO in the diferent presentations of it (single crystals, thin films, tapes, etc.)\cite{Chikumoto05,Mori10,Stangl25}.
Moreover, it aims to contribute to the development of scalable industrial processes, such as the production of YBCO-based superconductor tapes of second generation\cite{Wang24,LeeEngineeringSC}.

\section{Results}\label{SecResult}

\begin{figure}
\begin{center}
\includegraphics[width=0.9 \columnwidth]{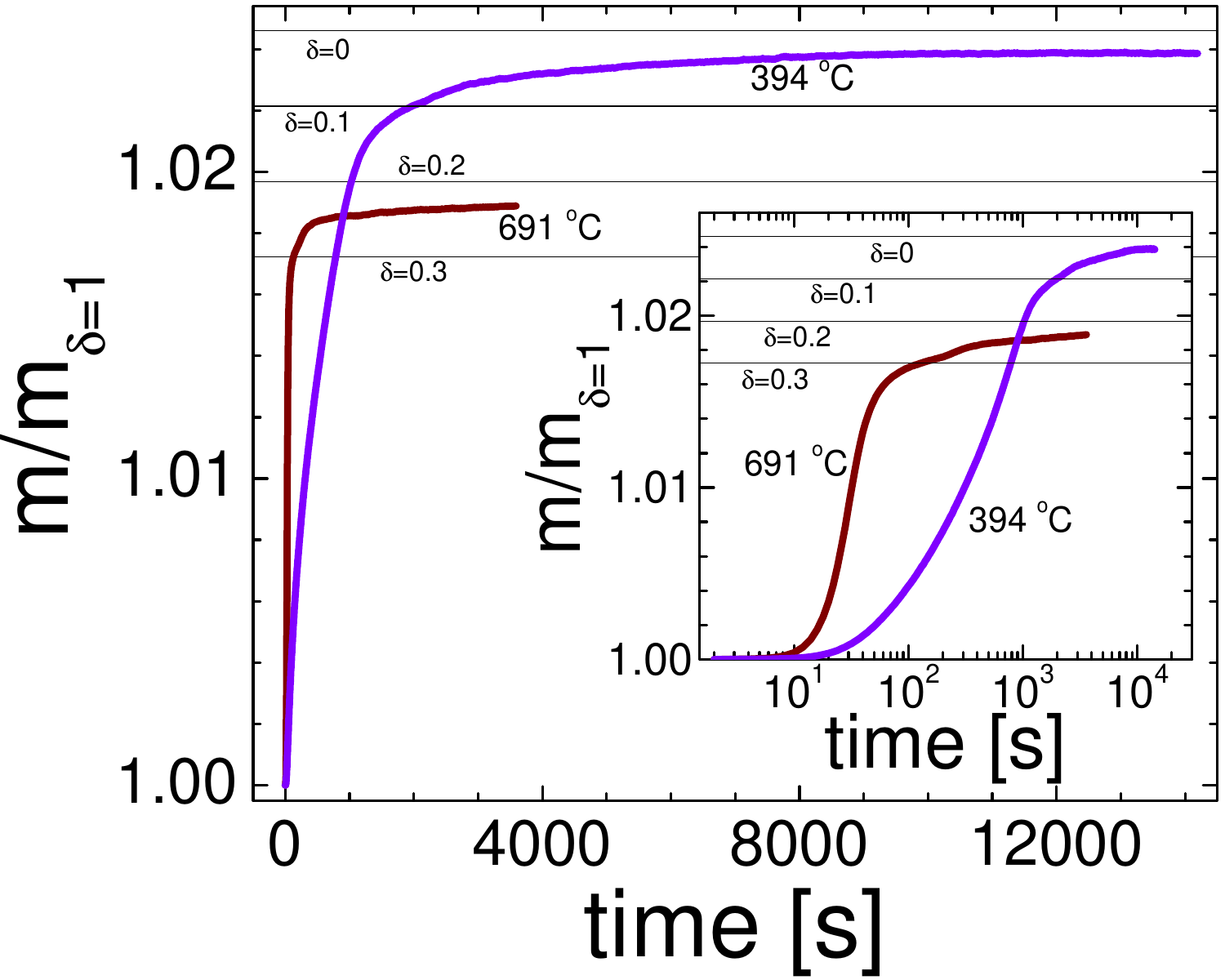}
\caption{\label{Fig0} 
Typical behavior of mass evolution of YBCO at constant temperature in an oxygen saturated atmosphere when starting from fully deoxygenated material.
Data is shown for two selected oxygenation temperatures: 394 $^o$C and 691 $^o$C.
Mass data is normalized by mass of same sample at fully deoxygenated condition ($\delta$ $=$ 1).
As a guide to the eye, horizontal black lines mark mass values for $\delta$ $=$ 0, 0.1, 0.2 and 0.3.
Inset shows same data with time (x-axis) in a logarithmic scale.
}
\end{center}
\end{figure}

Oxygen absorption was studied for YBCO powder with an average grain size of 5 $\mu$m obtained as described in Ref. \cite{LuccasArXiv}.
Mass variation measurements were performed under an oxygen atmosphere at constant temperature on samples of approximately 45 mg.
The evolution of the mass under these conditions was followed using a Shimadzu DTG-60H thermogravimetric balance, following the procedure detailed in Ref. \cite{LuccasArXiv}.
Starting from a fully deoxygenated material, the mass variation was measured under an oxygen atmosphere at constant temperature, for different temperatures in the range from 300 $^o$C to 800 $^o$C.
In general terms, the results are consistent with previous reports, showing that as the oxygenation temperature increases, the velocity for the oxygen absorption becomes higher, although the material reaches lower oxygen saturation levels; {\emph{i.e.}}, faster oxygen absorption at higher temperatures but leading to higher final $\delta$ values\cite{LuccasArXiv,Chen98,VazquezNavarro99,Qu13}.
Typical results of mass evolution during the oxygenation process are shown in Fig. \ref{Fig0} for two selected temperatures in the range under study.
Data are shown with time scale in both: linear (main) and logarithmic (inset) scale.
As a guide to the eye horizontal black lines are drawn at $\delta$ values of 0, 0.1, 0.2 and 0.3

\begin{figure}
\begin{center}
\includegraphics[width=0.9 \columnwidth]{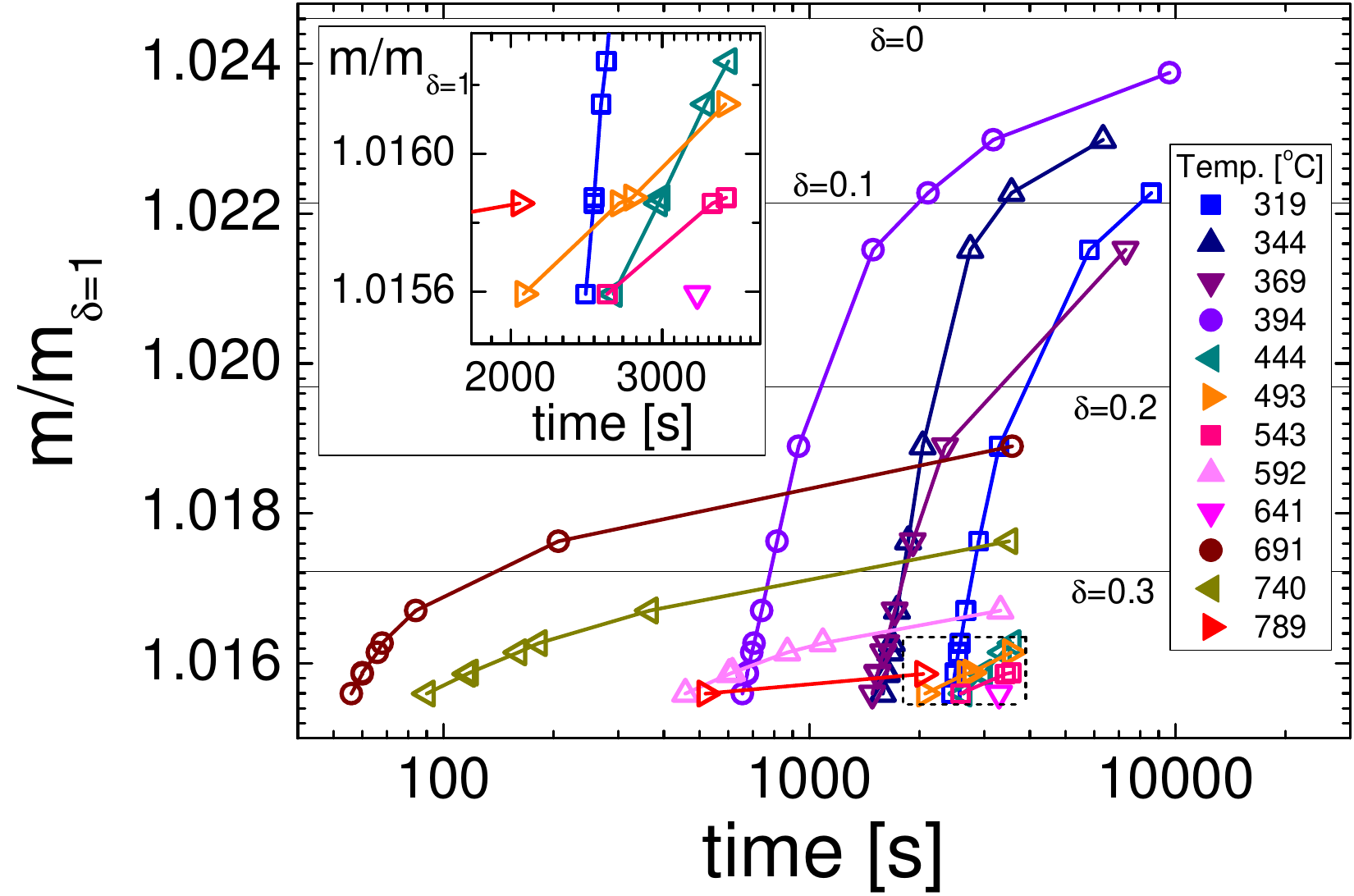}
\caption{\label{Fig1} 
Oxygenation of YBCO powder at constant temperature for different temperatures.
The figure shows the mass normalized to the fully deoxygenated mass value as a function of time.
The curves indicate the time required for the material to reach oxygen saturation value for each of the temperatures under study.
Every curve finished at the oxygen saturation value for their own temperature.
The intermediate points correspond to times at which the material reaches the oxygen saturation values obtained at the other temperatures under study.
The inset shows a detailed view of the region marked by a dotted rectangle in the main figure.
In general, it is observed that at higher temperatures the material reaches oxygen saturation faster, but at lower oxygenation levels (higher $\delta$ values).
}
\end{center}
\end{figure}

Based on the data obtained from oxygenation measurements at the different temperatures under study, particular attention was given to the behavior of the system near oxygen saturation.
Arbitrarily, only the mass values corresponding to oxygen saturation conditions at each temperature were taken as reference points.
In this way, a single reference mass value per temperature was defined, corresponding to the mass reached under saturation conditions at that specific temperature.
Then, for each temperature, the time required to reach these reference mass values was determined.
As a result, for temperatures at which the material saturates at lower oxygenation levels (lower mass, higher $\delta$ values), fewer points are available, since under these conditions the sample mass does not reach all the selected reference values.
This approach allowed a direct comparison of the time required, at each temperature, to reach the saturation $\delta$ values corresponding to the different temperatures under study.
The resulting analysis is summarized in Fig. \ref{Fig1}.

Fig. \ref{Fig1} shows, for each temperature under study, the time required for the material to reach its corresponding oxygen saturation value, as well as the time needed to attain the oxygen saturation values observed for the other temperatures.
It is noted that at 641 $^o$C only a single data point is present, since at this temperature the lowest oxygenation value (m/m$_{\delta = 1}$) was measured.
For the remaining temperatures, the final point of each curve (higher time value) corresponds to the oxygen saturation condition, while the intermediate points indicate the times required at each temperature to reach oxygen saturation levels associated to the other temperatures under study.

From this analysis, two temperature regimes are clearly distinguished.
In one hand, temperatures such as 691 $^o$C and 740 $^o$C, at which the material reaches high oxygenation levels within very short times.
On the other hand, temperatures such as 394 $^o$C, 344 $^o$C, 319 $^o$C, and 369 $^o$C, for which significantly higher oxygen saturation levels are achieved.
From these two groups, it is particularly noteworthy that oxygenation at 691 $^o$C enables rapid attainment of high oxygenation levels, while oxygenation at 394 $^o$C leads to the highest saturation levels.

\section{Discussion}\label{SecDisc}

The results presented in Fig. \ref{Fig1} are in good agreement with those reported in the literature, where the material exhibits oxygenation levels corresponding to $\delta$ $>$ 0.2 only for temperatures above 450 $^o$C\cite{Jorgensen87}.
These final oxygenation values correspond to theoretical critical temperature T$_C$ values significantly below the optimum, typically around 20 \% below the maximum achievable for the material (T$_C$ $\approx$ 70 - 80 $^o$C)\cite{Beyers89,Conder00,Chikumoto05}.
However, oxygen annealing processes are often reported at higher temperatures while still yielding higher T$_C$ values than theoretical ones\cite{Yu89,Zhang14,Cui05,Zheng03,Tu87}.
In addition, despite extensive research, there is still no established procedure or guide vector indicating the optimal movement direction in this area.
In other words, there is no a clear protocol for the oxygenation process for achieving the best superconducting properties of YBCO.

\begin{figure}
\begin{center}
\includegraphics[width=0.9 \columnwidth]{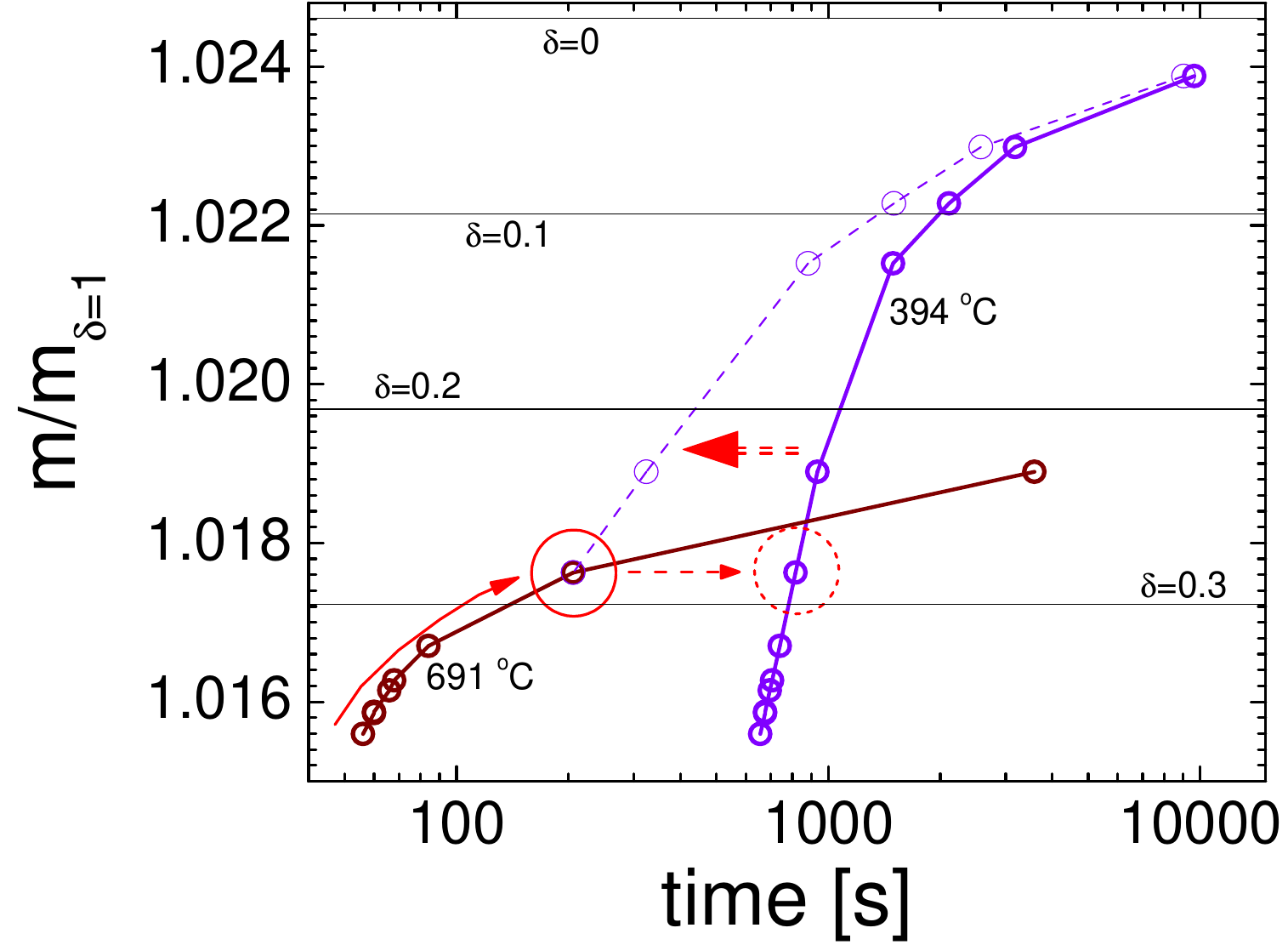}
\caption{\label{Fig2} 
Evolution of the mass of YBCO material as a function of time during the oxygenation process at constant temperature for 394 $^o$C and 691 $^o$C (continuous lines).
At higher temperature, a rapid increase in mass is observed, pointing out a fast oxygen absorption for the material.
At lower temperature, the process is slower but leads to higher oxygenation levels (lower $\delta$ values).
The dashed curve represents the proposed evolution of the material mass in a mixed-temperature oxygenation process, consisting of an initial oxygenation stage at high temperature (691 $^o$C) followed by a further oxygenation stage at lower temperature (394 $^o$C), with a transition between temperatures at the point indicated by a red dashed circle.
}
\end{center}
\end{figure}

In this context, and based on the results obtained in this work, particular attention is given to two oxygenation temperatures: 691 $^o$C and 394 $^o$C.
Fig. \ref{Fig2} shows exclusively the results corresponding to these two temperatures.

At 691 $^o$C, a very rapid oxygen absorption is observed during the initial stage, representing the fastest absorption rate among all the temperatures under study.
However, this rate progressively decreases, reaching saturation at $\delta$ $>$ 0.2.
In contrast, at 394 $^o$C, the oxygenation rate is lower but remains nearly constant, allowing the material to reach $\delta$ $<$ 0.2 and even to approach saturation values close to $\delta$ $\approx$ 0.

From these observations, it can be inferred that an optimal oxygenation process for YBCO should involve a combination of different temperatures.
Fig. \ref{Fig2} illustrates in dashed line the expected response of the material in an oxygenation process at 394 $^o$C when preceded by an initial oxygenation treatment at 691 $^o$C.
As indicated, an optimal process would consist of a first oxygenation step at 691 $^o$C for 210 s (3.5 min), followed by a second step of oxygenation at a temperature of 394 $^o$C, enabling the system to reach a near-optimal oxygenation ($\delta$ $\approx$ 0) while minimizing the total processing time.

\begin{figure}
\begin{center}
\includegraphics[width=0.9 \columnwidth]{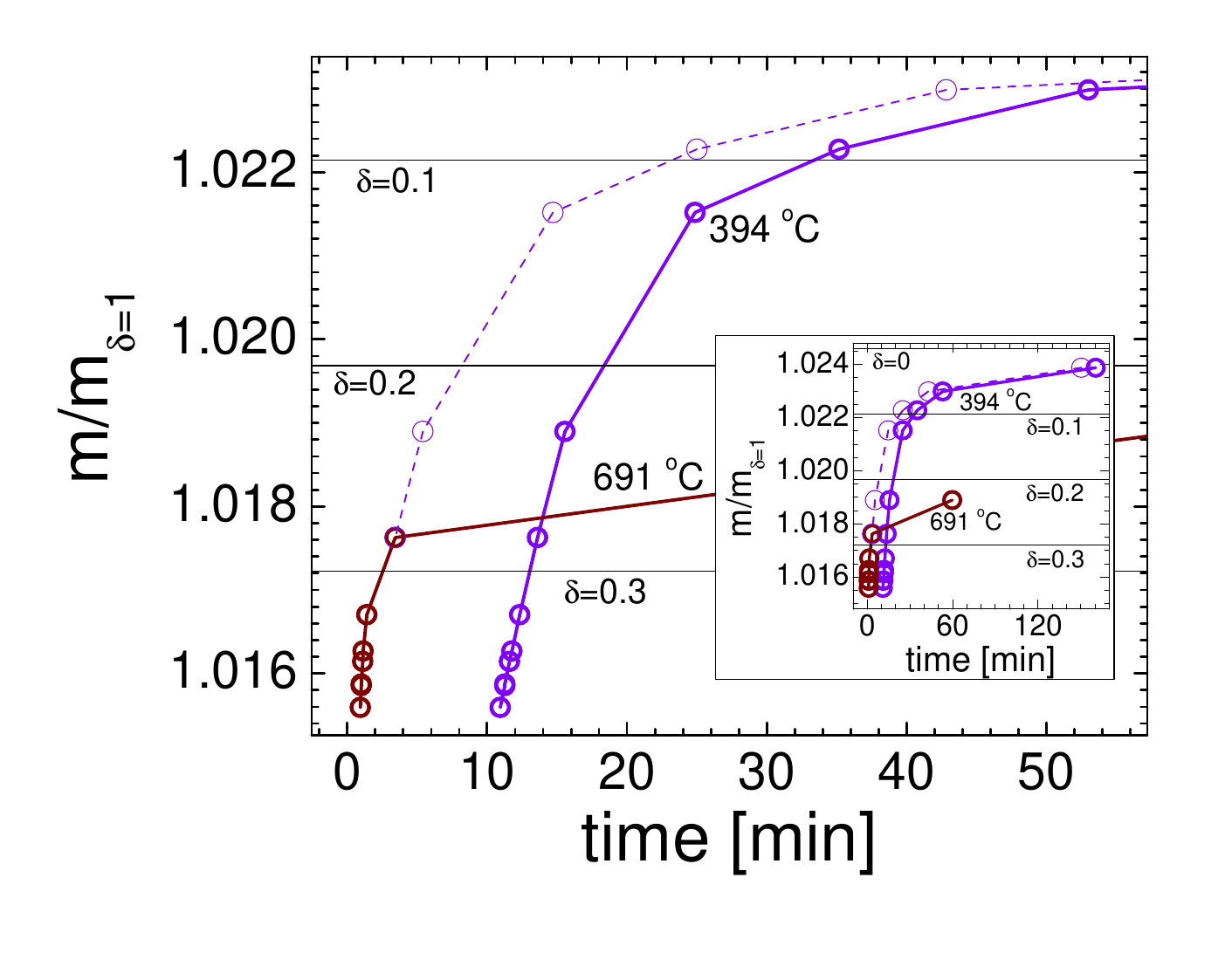}
\caption{\label{Fig3} 
Evolution of the mass of YBCO material as a function of time during the oxygenation process at constant temperature for 394 $^o$C and 691 $^o$C.
The figure presents the same data as in Fig. \ref{Fig2} with the time in the x-axis in linear scale.
It can be clearly observed that the temperature switch is proposed at the point where the slope of the mass evolution curve (directly associated with $\delta$) at high temperature becomes lower than that at low temperature.
In particular, the protocol proposed here allows to reach to a value of $\delta$ $\approx$ 0.12 in less than 60 \% of the time needed working only at 394 $^o$C.
The inset shows full data for both temperatures.
}
\end{center}
\end{figure}

Fig. \ref{Fig3} presents the same analysis showing time data in a linear scale.
Here, it can be clearly observed that, for the measured data, the time required for YBCO to reach $\delta$ $<$ 0.1 is reduced by approx. 30 \% when applying this temperature-combined strategy.

Different approaches to the oxygenation process can be found in the literature that are aimed at obtaining high superconducting performance of YBCO materials.
However, most of these methods rely on treatments at a single and high temperature\cite{Solovjov19,Yu89,Cui05,HudnerPhDThesis,Zheng03,Chikumoto05}.
Even when additional parameters such as pressure or growth conditions are varied, the oxygenation temperature is typically kept constant\cite{Zheng03,Chow92}.
In some cases, treatments involve annealing at high temperatures followed by cooling in an oxygen atmosphere\cite{Solovjov19,Chow92,Tu87,Chikumoto05,Mori10}, which can be considered an uncontrolled approximation to the simplest approach proposed in this work.
Notably, in such cases, cooling rates and temperature profiles during the process are generally not reported.

Given this diversity of reported approaches trying to achieve similar outcomes, the present study provides a systematic, as well as necessary, contribution to the field.
The work focuses on YBCO powder samples and demonstrates that optimization of the oxygenation process can be significantly improved by employing controlled temperature variations.
While absolute processing times may depend on the specific form of the material (bulk, powder, tapes, films, etc.), the optimal temperature conditions identified here are expected to remain relevant beyond that frontier.
Moreover, in this work we propose a protocol using the simplest oxygenation temperature combination, that is only two constant temperatures with a fix theoretical temperature change joining them.
However, even complexity could be increased with no control at all, a different combination of more than just two oxygenation temperatures could be proposed with even better results than the one exposed herein.
In essence, this study opens the possibility of exploring YBCO oxygenation process using variation temperatures protocols.

Results presented here, where oxygenation of YBCO is performed in powder samples, could be received as an exploration of a superficial behavior more than a bulk one.
However, and considering that thin/thick films of YBCO prepared in laboratories for studying properties of superconductor tapes are in thickness below the grain size of samples presented in this work\cite{Chikumoto05,Chow92,Chung07,Mori10,Villarejo21}, results show herein are highly relevant for applications, and allows using these assumptions as a basis for further optimization of processing times in industrial applications where brands work on these scales.

\section{Conclusions}\label{SecConcl}

In this work, the oxygenation behavior of powder of YBCO at constant temperature was investigated in the temperature range of 300 $^o$C to 800 $^o$C.
The evolution of the oxygenation $\delta$ as a function of time was followed through the mass variation measurements of the material, starting from completely deoxygenated samples.
$\delta$ values for saturation of oxygen was determined for each one of the working temperatures.
A comparative analysis between working temperatures of time necessary for reaching all the previous determined $\delta$ values was carried out.
Based on these results, a mixed-temperature oxygenation protocol is proposed, enabling optimization of the final oxygenation level while reducing the total processing time up to 60 \% in some particular conditions ($\delta$ $\approx$ 0.12 at constant oxygenation temperature of 394 $^o$C).

\section{CRediT authorship contribution statement}\label{SecCred}

RFL: Conceptualization, data curation, formal analysis, investigation, methodology, project administration, resources (YBCO powder samples), supervision, validation and visualization for all the stages involved in this research, as well as writing (original draft).
LG: Investigation (data measurements) and validation (data measurements).
Both authors contribute to reviewing this manuscript.

\backmatter

\bmhead{Acknowledgements}\label{SecTks}

Authors thank to J. A Malarr\'{\i}a for his scientific contribution.
This work was partially supported by MINCyT through the PICT-2017-2898 and PICT-2020-3758 FONCyT projects and by CONICET through the PIP-2020-2383 project.
This work was supported by the Argentinean scientific system against constant efforts of current Argentinean government\cite{TheyWantTheEnd}.

\bibliography{Luccas-EtAl_YBCOannealing_Refs}

\end{document}